\documentclass[10pt,sigplan,letterpaper,twocolumn]{article}
\usepackage[10pt,nocopyright]{sigmin}

\usepackage{algorithmic}
\usepackage{textcomp}
\usepackage{xcolor}
\usepackage{filecontents}
\usepackage{listings}
\usepackage{xcolor}
\usepackage{colortbl}
\usepackage{balance}
\usepackage{graphicx}
\usepackage[hyphens]{url}
\usepackage[hyphenbreaks]{breakurl}
\usepackage{hyperref}
\usepackage{threeparttable}
\usepackage[linesnumbered,boxed,lined,ruled,vlined]{algorithm2e}
\usepackage{setspace}
\usepackage{natbib}
\usepackage{diagbox}

\usepackage{subfig}
\usepackage{wrapfig}
\usepackage[export]{adjustbox}
\usepackage{times}
\usepackage{ifsym}
\usepackage{breqn}
\usepackage{bbding}
\usepackage{booktabs}
\usepackage{multirow}
\usepackage{amssymb}
\usepackage{enumitem} 

\usepackage{mathptmx}
\usepackage{mathtools}

\usepackage{mdframed}
\usepackage{mfirstuc} 

\usepackage{multicol}
\usepackage{rotating}
\usepackage{listings}
\usepackage[compact]{titlesec}
\titleformat*{\section}{\large\bfseries}
\titleformat*{\subsection}{\normalsize\bfseries}
\titleformat*{\subsubsection}{\normalsize\bfseries}
\titlespacing*{\section}{0pt}{*2}{*1}
\titlespacing*{\subsection}{0pt}{*1.5}{*0.8}

\usepackage{authblk}
\usepackage{longtable}
\usepackage{supertabular}
\usepackage{pifont}
\usepackage{threeparttable}
\usepackage{makecell}
\usepackage{soul}

\usepackage{tikz}

\newcommand{\sys}{\textsc{GIFT}\xspace}
\newcommand{\syscc}{\textsc{GIFT-CC}\xspace}
\newcommand{\cc}{confidential computing\xspace}
\newcommand{\sysmonitor}{GIFT monitor\xspace}

\newcommand{\tcb}{5.6K}

\newcommand{\methodone}{encryption-as-isolation\xspace}

\newcommand*\circled[1]{\tikz[baseline=(char.base)]{
            \node[shape=circle,draw,inner sep=0.1pt, minimum size=9pt] (char) {\footnotesize #1};}}

\newcommand{\batchedIFT}{Ownership Segmentation\xspace}
\newcommand{\whiteboxIFT}{Static Flow Analysis\xspace}
\newcommand{\asyncIFT}{Decoupled Flow Tracking\xspace}
\newcommand{\batchedift}{ownership segmentation\xspace}
\newcommand{\whiteboxift}{static flow analysis\xspace}
\newcommand{\asyncift}{decoupled flow tracking\xspace}

\newcommand{\stitle}[1]{\vspace{0.5ex}\noindent{\bf #1}}
\newcommand{\eul}[1]{\emph{\ul{#1}}}

\newcommand{\ccomment}[2]{{\bf[\textcolor{red}{#1}: \textcolor{blue}{#2}]}}
\newcommand{\TODO}[1]{{\ccomment{TODO}{#1}}}

\newenvironment{myitemize}%
  {\begin{list}{\labelitemi}{\itemsep1pt \topsep2pt \parsep0.00in
  \partopsep=0pt \leftmargin1.4em}}%
  {\end{list}}

\newenvironment{myenumerate2}
  {\begin{enumerate}[
    leftmargin=*,
    itemsep=0.2ex,
    topsep=2pt,
    parsep=0pt,
    partopsep=0pt,
    label=\arabic*),
    font=\normalfont]
  }
  {\end{enumerate}}

\newcommand{\pgwrapper}[3]{\begingroup \color{#1} #2: #3 \endgroup}

\newcommand{\cheng}[1]{\pgwrapper{blue}{cheng}{#1}}

\newcommand{\nkernels}{29}

\AtBeginDocument{%
  \providecommand\BibTeX{{%
    \normalfont B\kern-0.5em{\scshape i\kern-0.25em b}\kern-0.8em\TeX}}}

\begin{document}


\title{\Large \bf{
Here is a \sys: Enforcing User Data Isolation in LLM Serving via \\
\underline{G}PU \underline{I}nformation \underline{F}low \underline{T}racking
}}



\pagestyle{plain}

\author{
Jiacheng Shi$^{1}$, Xunjie Wang$^{1}$, Cheng Tan$^{2}$, Jinyu Gu$^{1}$\\
\small $^{1}$Institute of Parallel and Distributed Systems, School of Computer Science, Shanghai Jiao Tong University\\
\small $^{2}$Northeastern University
}

\date{}
\maketitle

\frenchspacing


\begin{abstract}
LLM serving frameworks process large volumes of user data---often containing
sensitive information---on shared infrastructure.
Ensuring isolation between users who share the same serving framework (on CPUs) and LLM operators (on GPUs)
is critical for privacy protection.

This paper presents \sys, a GPU Information Flow Tracking system
that enforces user data isolation in LLM serving with minimal overhead.
Moreover, the design of \sys is non-intrusive and allows CPU-side serving frameworks to evolve freely.
It rests on two key insights.
First, \emph{encryption-as-isolation} leverages the observation that CPU components only orchestrate data flow,
not content manipulation; thus, per-user encryption can provide isolation without modifying serving logic.
Second, GPU kernels exhibit limited and predictable information flows,
enabling \emph{\whiteboxift}.
\sys precomputes information flow rules for each kernel and
    uses \emph{\asyncift}, avoiding instrumentation or GPU stalls.

Furthermore,
we extend \sys to \syscc,
which integrates confidential computing to protect against untrusted operating systems and hypervisors (LLM service providers).
Implemented on vLLM and DistServe, \sys and \syscc enforce user
data isolation with a
4$\sim$10.7\% throughput overhead while maintaining the same latency level.

\end{abstract}

\section{Introduction}
\label{sec:intro}


Large Language Models (LLMs) have become increasingly integrated into users' daily activities.
For example, they assist with
drafting emails,
answering questions,
generating code,
and powering personal assistants across both
consumer and enterprise applications~\cite{chatgpt,deepseek,github-copilot,apple-intelligence,huawei-celia,microsoft-copilot}.
As these models handle sensitive tasks,
they increasingly observe and process private user data---ranging from personal
schedules and medical inquiries to proprietary business documents.

This growing need to handle sensitive data
raises fundamental concerns about
data leakage and information exposure control.
For example,
OpenAI temporarily took ChatGPT offline in 2023 due to a caching system bug
that allowed some users to see chat titles and histories from other active users~\cite{gpt-bug}.
More recently, in 2025, people found that ChatGPT content had appeared
in Google search results~\cite{gpt-searchable-google}.
%
%
These issues are not surprising given the rapid evolution of LLM serving systems---bugs
and vulnerabilities~\cite{cve-2025-32444,cve-2025-24357,cve-2025-47277,shelltorch,pytorch-rce,tensorflow-rce-eval,tensorflow-rce-yaml,triton-server-rce,transformers-rce-2023,transformers-rce-2024}
are inevitable, and they may lead to unintended data disclosure.

Indeed, an attacker can exploit vulnerabilities in LLM serving systems (e.g.,
vLLM)---such as CVE-2025-24357~\cite{cve-2025-24357}---to perform remote code execution and leak
sensitive user data, for example, the contents in the KV cache.
This new family of attacks (details in \S\ref{ss:ourattack})
enables the attacker to access other users' private conversations
without triggering traditional security mechanisms, as the data theft occurs \emph{entirely within GPUs}.
These incidents expose a critical question:
how can serving frameworks reliably enforce user data isolation while evolving rapidly?



The canonical approach to prevent information from propagating from a source to unintended outputs is
\emph{Information Flow Tracking} (IFT), also known as taint tracking~\cite{taint-analysis,taint-droid,data-lifetime,dytan,demand-emulation,lift,taint-enhanced-policy,taint-eraser}.
However, applying traditional IFT to LLM serving introduces two challenges:
\begin{myenumerate2}


    \item \emph{Complex and fast-evolving LLM systems.}
    Today's LLM serving frameworks are complex, implemented across multiple languages---typically a mix of Python, CUDA, and C++ or Rust.
    Moreover, they evolve rapidly, with new features and techniques emerging frequently.
    For example, vLLM~\cite{vllm} and SGLang~\cite{sglang} have 7.5K and 5K commits in the last year.
    Traditional IFT is difficult to implement in such sophisticated systems and struggles to remain compatible with rapidly evolving codebases.

    \item \emph{High runtime overhead.}
    Conventional GPU IFT incurs significant overhead~\cite{gpu-taint},
    which is incompatible with the high-throughput, low-latency requirements of production LLM serving.
    Ensuring efficient IFT without degrading model serving
    performance is a major obstacle.

\end{myenumerate2}



In this paper, we present \emph{GIFT} (\emph{GPU Information Flow Tracking}),
a new IFT design that tracks the flow of user data throughout LLM serving.
\sys is the first IFT system for LLM serving that achieves
strong isolation guarantees---(a) enforcing user data isolation,
(b) preserving the flexibility to evolve the serving framework,
and (c) achieving this with minimal performance impact.


GIFT builds on two key observations.
First, in LLM serving, the CPU does not manipulate the data directly---instead, the GPU
handles all data processing (data-plane)---while the CPU controls the computation by issuing GPU kernels (control-plane), as shown in Figure~\ref{fig:background}.
This separation means the CPU can operate on \emph{opaque data},
focusing only on placing the right data at the correct GPU memory locations.
Second, for serving workloads, GPU kernels are predefined and contain few branches.
As a result, their information flows can be analyzed ahead of time and tracked at runtime efficiently.

Based on the first observation, \sys adopts \emph{\methodone}.
In LLM serving, the CPU side executes complex and rapidly evolving
codebases---such as PyTorch (2.4M lines of code) and Transformers (615K)---that are difficult to secure and often contain many bugs and
vulnerabilities; for example, PyTorch has 18 reported CVEs and a number of vulnerabilities are identified in popular serving frameworks~\cite{liu2024demystifyingrcevulnerabilitiesllmintegrated}.
When compromised, an attacker could easily extract user data from CPU memory.
Hardening such frameworks is challenging due to their size and evolution rate.
\sys instead encrypts all user data with per-user keys
and stores only ciphertext in CPU memory.
A small, trusted component decrypts the data immediately before transferring it to the GPU.
This approach decouples security from complex frameworks, allowing
them to evolve while guaranteeing that any CPU-side data
exposure reveals only ciphertext.

Based on the second observation,
\sys introduces a new IFT design that combines \emph{\whiteboxift} and \emph{\asyncift}.
Traditional IFT mechanism instruments programs to track data propagation,
but this is prohibitively expensive in LLM serving,
where GPU kernels are highly optimized and performance-critical.
Instead, \sys analyzes each GPU kernel offline to identify its possible information flows
and derive corresponding \emph{IFT rules}---deterministic logic dictating how the flow tracking metadata should be updated for that kernel.
During runtime, \sys executes these rules on the CPU in a decoupled manner
and then detects unauthorized data flows
without instrumenting the executing kernel or incurring instruction-level overhead on GPU.

\sys's design is general and applicable to various serving frameworks.
To demonstrate this, we implement \sys on two popular LLM serving systems:
vLLM~\cite{vllm} and DistServe~\cite{dist-serve}.
Furthermore, we extend \sys to operate under a stronger threat model where the underlying system software,
including the OS and hypervisor, is untrusted.
This extension, called \syscc, leverages hardware confidential computing
and is also implemented on both vLLM and DistServe,
which can isolate user data from the service providers.

Our evaluation across multiple LLM models shows that \sys adds negligible
latency overhead and incurs at most a 5\% throughput reduction under fully
saturated serving workloads.
\syscc, which integrates confidential computing hardware,
introduces up to a 7\% throughput overhead compared to non-confidential baselines.
In experiments with speculative decoding, \syscc experiences up to a 10.7\% throughput drop,
largely due to the additional CPU--GPU communication overhead introduced by
NVIDIA's confidential computing interface.

In summary, we make the following contributions: 1) \emph{\sys}: The first user data flow tracking approach designed for LLM serving systems;
2) \emph{\syscc}: An extension of \sys that leverages confidential computing for a stronger threat model;
3) Real implementation and extensive evaluation demonstrate negligible performance overhead.

\section{Motivation, Challenges, and Threat Models}
\label{sec:motivation}


\subsection{LLM Serving Systems}
\label{sec:bg-llm-inference-service}

\begin{figure}[tbp]
  \centering
  \includegraphics[width=0.85\columnwidth]{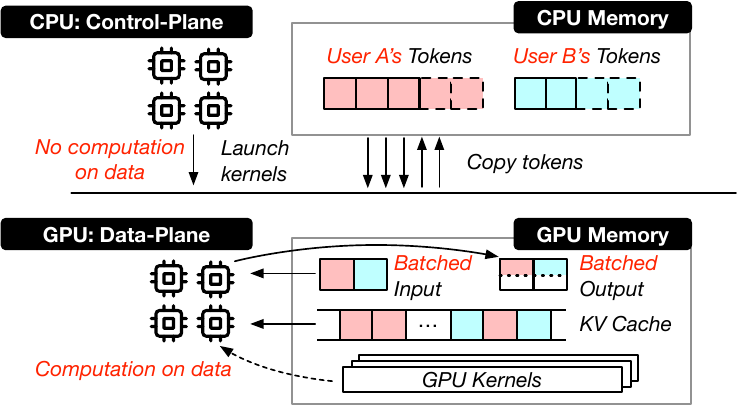}
  \caption{Control/data plane separation in serving systems.}
  \label{fig:background}
\end{figure}

\stitle{Data residency.}
LLM serving systems, such as vLLM~\cite{paged-attention} and DistServe~\cite{dist-serve},
orchestrate interactions between CPUs and GPUs to deliver model inference.
In these systems, CPUs primarily handle
\emph{control-plane} operations such as scheduling, batching, and issuing GPU kernels.
The GPUs then execute these kernels to perform computation-intensive tasks (\emph{data-plane} operations) like attention and matrix multiplication.
Importantly, CPUs typically do not modify or inspect the content of user data; they only
coordinate the execution flow.
As a result, most private or user-specific
data---including intermediate activations and KV caches used for
context reuse---reside on GPUs during serving.
Figure~\ref{fig:background} overviews this separation.

\stitle{System evolution and complexity.}
With the rapid growth of AI,
LLM serving systems have evolved---and continue to evolve---into
complex infrastructures incorporating numerous optimizations~\cite{paged-attention,orca,splitwise,dist-serve,sarathi-serve,speculative-decoding,infini-gen,loong-serve}.
This complexity arises for good reasons---improving GPU utilization~\cite{splitwise, dist-serve},
optimizing throughput and latency~\cite{sarathi-serve},
and enabling dynamic batching~\cite{orca}, caching~\cite{paged-attention}, and speculative decoding~\cite{speculative-decoding}.
However,
these same optimizations introduce more sophisticated interactions between components,
making reasoning about data privacy increasingly difficult.

\subsection{Data Privacy Threats in LLM Serving Systems}
\label{ss:data-privacy-threats}

Data privacy is fundamental to LLM services (e.g., ChatGPT, Google Gemini, Claude Code),
as user prompts often contain sensitive information~\cite{Yao_2024,yan2024protectingdataprivacylarge}.
Prior cloud incidents~\cite{log4shell,snowflake-2024-breach,toyota-2023-breach,capitalone-2019-breach} 
have shown that privacy leaks can arise from diverse sources---bugs, vulnerabilities, misconfigurations, operator errors, and attacks.
Moreover, the risk increases as LLM serving frameworks evolve
rapidly, incorporating larger and increasingly complex codebases.

We conduct a survey of five popular AI infrastructure systems---PyTorch~\cite{pytorch}, Transformers~\cite{transformers},
ONNX~\cite{onnx}, TensorFlow~\cite{tensorflow}, and Triton Server~\cite{triton-server}---examining their codebase sizes,
reported vulnerabilities (i.e., CVEs),
and proof-of-concept exploits.
Table~\ref{tab:ai-frameworks} summarizes the results.
Each framework or system typically has a large codebase and
exploitable vulnerabilities.

\begin{table}[htb]
    \fontsize{9pt}{9pt}\selectfont
    \centering
    \caption{Code sizes, CVEs, and proof-of-concept exploits}
    \label{tab:ai-frameworks}
    \begin{tabular}{l|ccc}
      \toprule
      \textbf{Component} &\textbf{Lines of Code}  &\textbf{\#CVE}  &\textbf{Proof-of-Concept} \\ 
      \midrule
      vLLM &146K &31 &\cite{vllm-poc-1}, \cite{vllm-poc-2}, \cite{vllm-poc-3} \\
      PyTorch &2.4M &18 &\cite{shelltorch}, \cite{pytorch-rce}, \cite{pytorch-file-write} \\
      Transformers &615K &4 &\cite{transformers-rce-2023}, \cite{transformers-rce-2024} \\
      ONNX &128K &4 &\cite{onnx-file-write}, \cite{onnx-dir-traversal} \\
      TensorFlow &1.9M &439 &\cite{tensorflow-rce-eval}, \cite{tensorflow-rce-yaml}, \cite{tensorflow-data-leak} \\
      Triton Server &78K &7 &\cite{triton-server-rce} \\
      \bottomrule
    \end{tabular}
\end{table}

\label{ss:gptbug}
Beyond theoretical risks, privacy leaks have occurred in practice.
In 2023, ChatGPT experienced a major cross-user privacy leak
caused by a bug in its serving system~\cite{gpt-bug}.
During the incident, some users reported seeing chat histories of others,
resulting in OpenAI taking ChatGPT offline temporarily to apply a fix.

\stitle{GPU confused-deputy attacks.}
\label{ss:ourattack}
Unlike traditional attacks, those targeting LLM serving systems can exploit
GPU-level leakage channels that existing techniques are not able to handle.
To illustrate their implications, we conduct an example attack---
the compromised serving framework directs the GPU to execute
benign kernels with maliciously crafted parameters, which is similar to the confused-deputy attacks~\cite{confused-deputy}.

Specifically, vLLM adopts the paged attention~\cite{paged-attention} mechanism,
using a \texttt{SelfAttnBlockSpaceManager} object to manage the allocation of KV blocks among different requests.
The object provides a method \texttt{get\_block\_table} that returns the KV blocks allocated for a certain request ID.
We exploit one remote-code-execution (RCE) vulnerability CVE-2025-24357~\cite{cve-2025-24357} in vLLM (before v0.8.0) and thus replace \texttt{get\_block\_table} with a malicious one that always returns the KV blocks of the first request.
Thus, the K/V vectors of the first request will always be used to
calculate the attention scores of the other requests, 
leaking one user's data into the responses of other users.

We verify this attack by emulating two users, Alice and Bob, who send prompts ``Hello, I am Alice/Bob.'' to the vLLM server simultaneously.
Both users receive the same response: ``Hello, Alice! \dots'', meaning that Alice's name is leaked to Bob.

\stitle{User data isolation.}
User data isolation is a fundamental (yet often implicit) assumption in LLM serving systems.
Users expect that their interactions---queries, prompts, and responses---remain private and separated from those of others.
This expectation follows naturally from how models are served:
each user session is independent, and there is no legitimate need for data belonging to
different users to be mixed or shared.
In practice, however, ensuring user data isolation is nontrivial as discussed above.

\subsection{Ensuring User Data Isolation via IFT}
\label{ss:useriso}

Information Flow Tracking (IFT)~\cite{taint-analysis,taint-droid,data-lifetime,dytan,demand-emulation,lift,taint-enhanced-policy,taint-eraser} 
provides a principled mechanism to enforce isolation by monitoring how data propagates within a system.
However, applying classic IFT directly to LLM serving systems fails due to two fundamental challenges:
\begin{myenumerate2}

\item On the CPU side, the serving frameworks are complex systems with multiple languages and many third-party libraries.
For example, vLLM uses C++ and Python, and imports 244 dynamic libraries.
The frameworks also evolve rapidly: vLLM has 7.5K commits in the last year.
Implementing IFT logic for each framework version requires high engineering effort.
Moreover, there are no mature open-source multi-language dynamic IFT tools, and
research work~\cite{multi-lang-taint} shows instrumentation can cause up to {2.6\texttimes} overhead on Python programs.

\item On the GPU side, traditional instruction-level instrumentation 
introduces high 
runtime overhead.
Modern LLM serving is extremely sensitive to latency,
so practitioners will not adopt GPU kernels
that run {3\texttimes} slower~\cite{gpu-taint}, 
regardless of the isolation guarantees they provide.

\end{myenumerate2}


A practical IFT approach for LLM serving must therefore satisfy two criteria:
(a) it should integrate easily with existing serving systems without extensive re-engineering, and
(b) it should impose minimal performance overhead.

\sys is \emph{the first} system satisfying these criteria.
It provides user data isolation 
with negligible performance overhead: no observable latency increase and only a 4\% throughput reduction.
\sys can adapt to diverse frameworks 
and hardware configurations; 
to demonstrate this,
we implement \sys on two serving frameworks: vLLM and DistServe.
Orthogonal to the choice of framework, we also build two variants---with and
without \cc (NVIDIA-CC), a hardware-based security feature~\cite{nvidia-cc}. 
In the following, we use \syscc to denote the versions (for both
vLLM and DistServe) with \cc enabled, and \sys to denote
those without.

\begin{figure*}[ht]
  \centering
  \includegraphics[width=\textwidth]{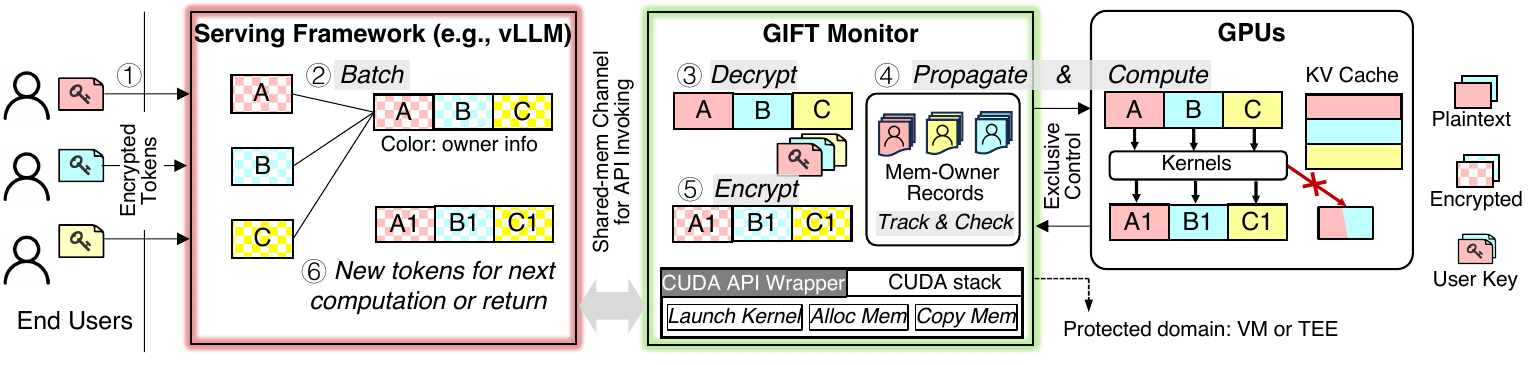}
  \caption{System overview: workflow and architecture.}
  \label{fig:system-overview}
\end{figure*}

\subsection{Threat Models}
\label{subsec:threat-model}

\stitle{Threat model for \sys.}
\sys targets a setting where external attackers and buggy code represent the primary threats.
\sys trusts the underlying operating system and hypervisor, including their enforcement of process and VM isolation.
In contrast, the LLM serving framework is treated as untrusted---it may be
compromised or exhibit arbitrarily malicious behavior
including redirecting one user's computation to access another's data (as demonstrated by our example attack, \S\ref{ss:ourattack}).
To enforce isolation guarantees, \sys relies on a trusted
component, \emph{\sysmonitor} (described in \S\ref{sec:overview}), which runs as a standalone VM
with 5.6K lines of code implemented in Rust.

\stitle{Threat model for \syscc.}
\syscc builds on the same core threat assumptions as \sys:
it considers LLM serving frameworks to be exploitable, while trusting the \sysmonitor for enforcement.
However, unlike \sys, \syscc does not trust the operating system or hypervisor.
Instead, it leverages \cc to ensure the confidentiality of user data even
in the presence of a compromised host.
In this model, adversaries---such as a privileged administrator of an outsourced
LLM service in the cloud (collecting user data for training)---may attempt to inspect CPU memory,
issue DMA operations to read GPU memory,
or conduct physical attacks like cold booting~\cite{cold-boot} to break memory confidentiality.

\stitle{Out-of-scope attacks for \sys and \syscc.}
We do not consider side-channel attacks in \sys or \syscc, as such attacks are
orthogonal concerns that can be mitigated using complementary defenses~\cite{pcc-padding}. 
Similarly, we exclude denial-of-service (DoS) attacks
and rely on dedicated DoS detection and mitigation systems to address them.
Finally, we do not address issues related to the integrity of inference results,
assuming that users can detect significantly degraded or incorrect outputs.


\section{\sys Overview}
\label{sec:overview}



In \sys, each user has ownership of their data,
including input prompts and
all derived data generated during serving, such as KV caches, intermediate tokens, and output data.
Each user's private data is encrypted with a user-specific symmetric encryption key,
accessible only to the user and the trusted \sysmonitor.

\sys performs comprehensive ownership tracking%
\footnote{Ownership is equivalent to labels in classic IFT.
We use the term ownership to emphasize that GIFT enforces user data isolation
rather than the traditional separation between sensitive and non-sensitive data.}
throughout the entire data
lifecycle, maintaining strict isolation boundaries between users.

\stitle{System architecture.}
As shown in Figure~\ref{fig:system-overview},
\sys deploys a trusted component (\sysmonitor) within a protected domain to exclusively operate the GPUs,
while the untrusted LLM serving framework (e.g., vLLM) runs outside the domain.
When the operating system and hypervisor are trusted, 
\sys relies on them to allocate GPUs exclusively to the \sysmonitor,
and the serving framework and {\sysmonitor} can be isolated in two VMs.
In contrast, when they are untrusted, 
{\sysmonitor} can also be protected with a trusted execution environment (TEE) to implement {\syscc},
which will be introduced in \S\ref{sec:cc}.

\stitle{System workflow.}
At a high level, \sys operates as follows.

\circled{0} During system initialization, {\sysmonitor} registers all GPU kernels required for executing the target models.
In the later serving phase, it acts as the kernel launcher, permitting only registered kernels and
rejecting any unauthorized ones. It also loads model parameters into GPU memory and marks them as ``parameter''-owned.
Besides, it also stores user-specific keys into its protected memory.

\circled{1}
Tokenization of plaintext prompts, which is a lightweight computation,
occurs on end users' devices to maintain privacy.
The tokenized prompts are encrypted with per-user keys before being sent to the LLM serving framework in the cloud.

\circled{2}
The serving framework (e.g., vLLM) operates entirely on encrypted data while
handling control-plane tasks such as continuous batching and kernel scheduling.
When launching GPU kernels, it sends a request to \sysmonitor that includes
ownership metadata for the batched encrypted inputs.

\circled{3}
\sysmonitor decrypts user data using the corresponding owners' keys.
Crucially, even if the serving framework provides manipulated ownership metadata, the result is
only a decryption failure, not a privacy violation.

\circled{4}
After decryption, \sysmonitor launches the designated GPU kernels, transfers the plaintext data
to GPU memory, and records the ownership of GPU memory regions for the kernel inputs.
For GPU computation, it tracks and updates GPU memory ownership to serve two functions:
\begin{itemize}
    \item \sysmonitor needs to enforce strict user data isolation by preventing the framework
        from mixing data across users (e.g., adding one user's data to another's result) 
        or conducting confused-deputy attacks (\S\ref{ss:ourattack}).
        This mechanism preserves a fundamental invariant in LLM serving:
        throughout batched execution, each memory byte containing private
        information must have a singular ownership.

    \item Before returning data back to the untrusted framework (\circled{6}), 
          \sysmonitor needs to ensure that all data are encrypted using their corresponding owners' keys (\circled{5}). 
          Thereby, the framework always only sees encrypted data, whereas
          all the plaintext data only resides within the protected domain including the GPUs.
\end{itemize}

\stitle{Encryption-as-Isolation.}
\sys adopts \emph{encryption-as-isolation} as a core design principle to achieve user
data isolation while remaining compatible with the rapid evolution of LLM
serving frameworks.
The key insight is that encryption can serve as a
mechanism for isolation: even if a buggy or malicious framework manipulates
ciphertexts, it cannot extract meaningful information or mix data across users,
since each user's data is encrypted with a distinct key.

This approach builds on a general system idea---using encryption boundaries to
decouple different information flows\footnote{IFT systems often require declassification rules to determine which data should be released to lower security levels and how~\cite{Declassification}. In {\sys}, we use encryption to allow the flow of sensitive data.}, typically between trusted and untrusted components~\cite{cipherbase,stealthdb,maintainable-edb}.
Fundamentally, encryption-as-isolation works for \sys
because the CPU-side framework does not
modify the semantic content of data; it only orchestrates computation and data movement.
From the framework's perspective, ciphertexts and plaintexts are
indistinguishable, enabling continuous framework evolution without compromising
privacy or requiring changes to \sys's logic.

\stitle{A major challenge: IFT overhead.}
A key challenge in \sys lies in how ownership is propagated and checked during GPU computation.
We have so far described the conceptual workflow without detailing this
step, as efficient ownership tracking remains one of the most difficult
problems in achieving practical IFT for LLM serving.

Applying traditional IFT techniques directly to GPUs---where each byte or cache
line is labeled with ownership information and every instruction is
instrumented to propagate these labels---would introduce prohibitive overhead.
Such fine-grained tracking is fundamentally incompatible with the performance
demands of modern LLM serving.
In the following section, we present our insights and solutions to 
enable efficient GPU memory ownership tracking with near-zero overhead.

%


\section{Near-Zero-Overhead GPU IFT}
\label{sec:design}



The end-to-end performance overhead of Information Flow Tracking (IFT) depends on three key factors:
(a) the volume of data elements being tracked,
(b) the computational complexity of ownership tracking methods, and
(c) the runtime effects on the original GPU computation.
For LLM serving, such overhead must be near zero---otherwise, no practitioner
would adopt it in latency-critical production systems.

To meet this requirement, we systematically analyze the GPU
kernels used in LLM serving (\S\ref{ss:insights})
and develop a new IFT design (\S\ref{ss:batch}--\ref{ss:async})
that leverages data flow patterns of today's LLM GPU kernels
to achieve near-zero-overhead GPU IFT.
We begin by presenting our key insights.

\subsection{Observations and Insights}
\label{ss:insights}

We study the three factors of IFT overhead---data volume,
tracking method complexity, and runtime effects on GPU---and design corresponding techniques to
reduce each cost for LLM serving on GPUs.

\stitle{Insight 1: \batchedIFT.}
We observe that during LLM serving, data elements belonging to the same owner
typically cluster into contiguous GPU memory segments.
This spatial locality enables us to implement ownership recording
at the segment level rather than the byte or element level,
dramatically reducing metadata overhead and complexity.

\begin{figure}[htbp]
  \centering
  \includegraphics[width=1\columnwidth]{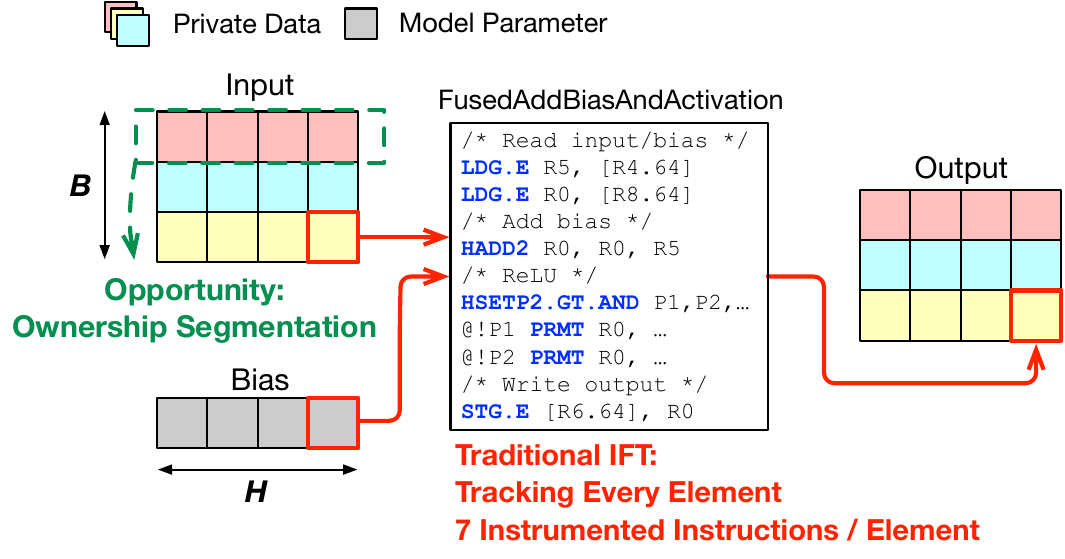}
  \caption{A kernel example and the opportunity for \batchedift. $B$ and $H$ indicate batch size and hidden size.
  }
  \label{fig:coarse-grained}
\end{figure}

Figure~\ref{fig:coarse-grained} illustrates this principle with a
representative feed-forward network kernel that adds a bias vector to each row
of an input matrix, computes ReLU, and stores the results in an output matrix.
While an element-level method would require maintaining ownership records for each
individual element---amounting to $B \times H$ discrete ownership
records---our batched approach requires only $B$ records for the entire input or
output matrix.

\stitle{Insight 2: \whiteboxIFT.}
Traditional IFT techniques~\cite{taint-analysis,taint-droid} are designed for CPU programs that exhibit
dynamic semantics with many branches, significant nondeterminism from various sources,
and often lack source code access.
In contrast,
LLM GPU kernels are typically deterministic,
have limited control-flow branching,
and are available in source form. 
These properties enable static reasoning and white-box analysis ahead of execution.
So, instead of instrumenting binary instructions,
we analyze kernel behavior offline and generate IFT rules that define ownership
propagation based on kernel arguments and their corresponding control-flow
branches, which are typically few in number.

For the kernel example in Figure~\ref{fig:coarse-grained}, conventional
instrumentation would necessitate instrumenting 7 instructions for propagating ownership. 
Owing to static analysis, \sys condenses this to just three streamlined operations required during runtime:
(1) verifying the bias vector as model parameters, (2) reading the input row
ownership records, and (3) propagating ownership to corresponding output rows.
This optimization reduces the computational complexity from $7 \times B \times
H$ operations to merely $2 \times B + 1$.

\stitle{Insight 3: \asyncIFT.}
By leveraging \whiteboxift,
the ownership propagation is no longer intertwined with the GPU kernel execution.
This enables us to decouple ownership tracking and offload it to CPUs 
(also owing to the reduced computational complexity),
effectively hiding the IFT overhead behind GPU computation.

\begin{figure}[htb]
  \centering
  \includegraphics[width=0.92\columnwidth]{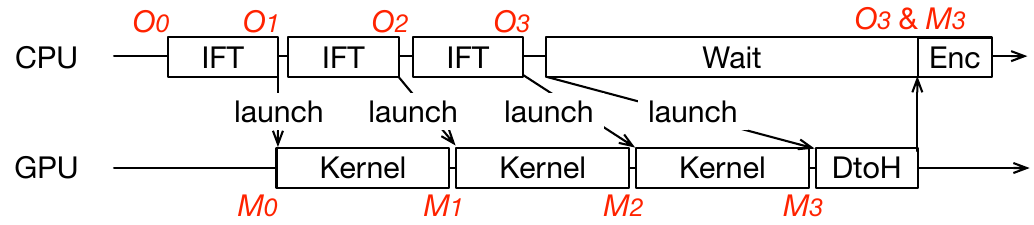}
  \caption{Overlapping IFT on CPU with GPU kernels. $O$: ownership state, $M$: GPU memory state, $O_i$ aligns with $M_i$, DtoH: device-to-host memory copying, Enc: encryption.}
  \label{fig:asynchronous}
\end{figure}

\stitle{Malicious behavior detection during tracking.}
Before launching a kernel as requested by the LLM serving framework,
\sysmonitor performs ownership tracking according to its IFT rule and the provisioned kernel arguments.
If there is no unexpected behavior such as ownership mixing, the kernel will be launched;
otherwise, \sysmonitor will reject the kernel launch request.

As shown in Figure~\ref{fig:asynchronous}, \sysmonitor always performs ownership tracking 
before launching kernels and proceeds with subsequent IFT operations without waiting for kernel completion.
While this parallelization creates temporary misalignment between the logical
ownership state ($O$) and the actual GPU memory state ($M$), correctness is
maintained for two reasons: First, the ownership state seen by an IFT operation (e.g., $O_1$) 
always aligns with the memory state seen by the corresponding kernel (e.g., $M_1$),
because CPU IFT and GPU kernels are both executed sequentially.
Second, when data is exported from GPU,
the memory state ($M_3$) always aligns with the ownership state ($O_3$) used for key selection,
because we explicitly wait for the device-to-host memory copying kernels.

Next, we introduce each of the three techniques in detail.

\subsection{\batchedIFT}
\label{ss:batch}
\label{sec:ownership-record}

\sysmonitor maintains ownership records for each GPU.
These records---stored within the \sysmonitor protected domain---indicate readable GPU
memory regions and their corresponding data owners.

\stitle{Recording granularity.}
Each GPU memory block allocated via \texttt{cudaMalloc} contains an $n$-dimensional tensor that may store data from multiple users. 
Data belonging to the same user typically forms segments in these tensors, enabling coarse-grained ownership recording.

Figure~\ref{fig:coarse-grained} already shows a concrete example. 
As another typical example, PagedAttention~\cite{paged-attention}
partitions a 5-dimensional tensor $C$ into KV blocks for different users.
Each block consists of multiple K/V vectors, allowing ownership recording at KV
block level through tensor slices (e.g., $C$[1,2,:,0:8,:]).

\begin{figure}[htbp]
  \centering
  \includegraphics[width=1.0\columnwidth]{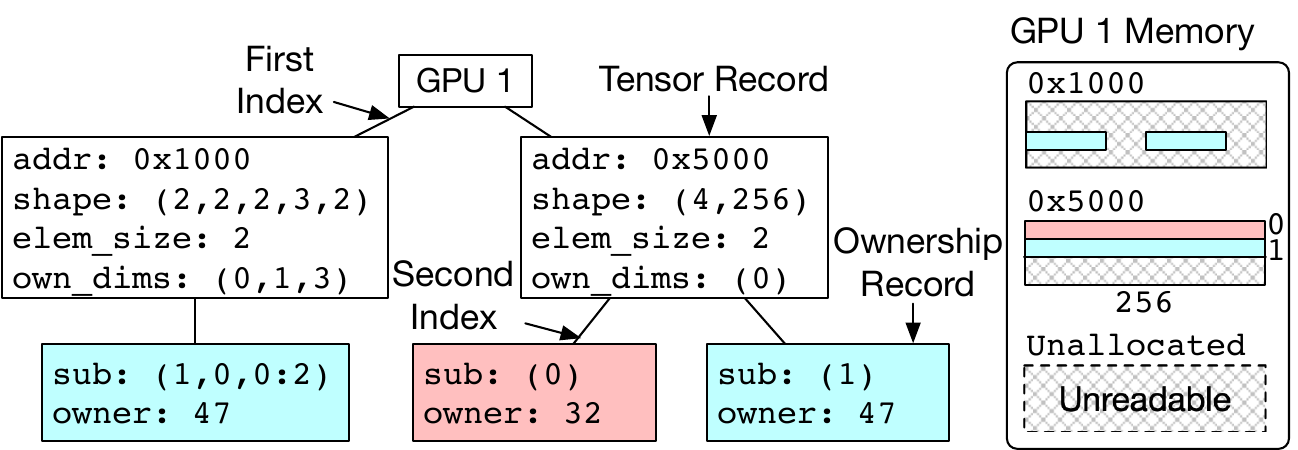}
  \caption{The data structure of ownership records, addr: base address, dtype: data type, odim: ownership-related dimensions, sub: subscript values.}
  \label{fig:ownership-record}
\end{figure}

\stitle{Ownership data structure.}
As illustrated in Figure~\ref{fig:ownership-record}, for storing ownership,
\sys uses two-level indexing: the first-level index maps base addresses to tensor records,
while the second-level index maps subscript values to ownership records within each tensor.

A tensor record consists of four fields: the base address, tensor shape, element size, and ownership-related dimensions. The last field specifies which dimensions determine data ownership, such as (0) for row number in a matrix (e.g., a hidden state matrix) or (0,1,3) for a 5-dimensional tensor (e.g., KV cache blocks).

Each ownership record contains two fields: (1) the subscript values of the ownership-related dimensions, and (2) the owner's user ID, or a special ID for model parameters.

\stitle{Unowned memory.}
To prevent leakage of legacy private data, \sysmonitor prohibits reads to any GPU memory region not covered by their ownership records,
and clears all ownership records of a tensor when releasing it.

\subsection{{\whiteboxIFT}}
\label{ss:aot}
\label{sec:kernel-ift}

The generation of \emph{IFT rules}---program routines that describe how ownership
information propagates from kernel inputs to outputs---includes three steps:
(1) extracting each kernel's \emph{Information Flow Graph (IFG)}, 
which captures data dependencies and transformation paths;
(2) validating the IFGs against the kernel implementations via symbolic execution~\cite{klee} 
to ensure consistency;
and (3) generating IFT rules based on the validated IFGs to define expected ownership propagation behavior.

\begin{figure}[htbp]
  \centering
  \includegraphics[width=0.95\columnwidth]{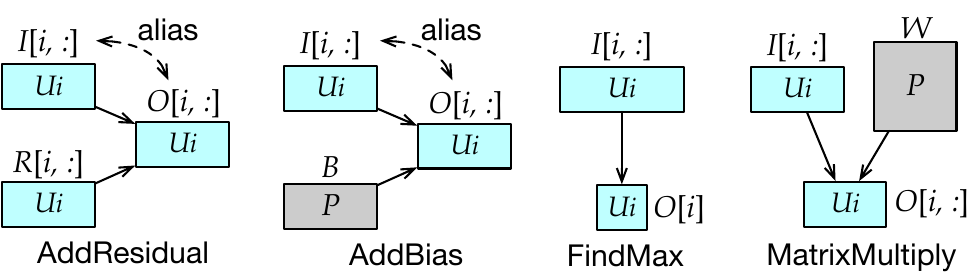}
  \caption{IFGs of four example kernels. $I$: input, $O$: output, $R$: residual, $B$: bias, $W$: weight, $P$: parameter, $U_i$: owned by user $U_i$. Alias: one node is an alias of another node.}
  \label{fig:coarse-grained-ifg}
\end{figure}

\stitle{Building IFGs.}
Figure~\ref{fig:coarse-grained-ifg} illustrates the IFGs for four example kernels, where each node represents a GPU memory segment owned by a single user. 
\texttt{AddResidual} adds a residual matrix to an input matrix. 
\texttt{AddBias} adds a bias vector to each row of an input matrix. 
\texttt{FindMax} finds the index of the maximum element in each row of an input matrix.
\texttt{MatrixMultiply} multiplies an input matrix by a weight matrix.
These IFGs are manually constructed by examining the information flow of kernels 
and the ownership layout of input tensors for benign inference.
Each node is represented by the tensor name, subscript values, and owner. The owner is a variable associated with the subscript values.

%

\stitle{Validating IFGs.}
\sys uses symbolic execution~\cite{symbolic-exec}
to validate that the manually constructed IFGs accurately
reflect the kernel implementation.
It first transforms each CUDA kernel into an equivalent C++ version compatible with the
widely used symbolic execution engine KLEE~\cite{klee}.
It then verifies that the kernel produces no information flows beyond those specified in its IFG.

\stitle{Generating IFT rules.}
Given a validated IFG, the program of its IFT rule should consist of the following steps:
(1) identifying tensor records corresponding to accessed tensors by matching their base addresses to kernel arguments;
(2) verifying that the kernel arguments are consistent with the expected tensor shapes;
(3) checking that tensors occupy distinct memory regions unless explicitly intended to alias, thereby detecting unintended aliasing;
(4) ensuring that parameter nodes are correctly labeled as parameter-owned in their ownership records;
(5) finally, iterating over all possible subscript values (e.g., all possible $i$ in Figure~\ref{fig:coarse-grained-ifg}),
    confirming that non-parameter source nodes share the same owner and assigning this ownership to the corresponding destination nodes.

\label{ss:semiauto}
\stitle{Semi-automated process and future work.}
Currently, IFT rule generation in \sys is semi-automated and requires developer assistance.
Although not yet fully automated, this process is a one-time effort per kernel.
Moreover, IFT rules are reusable and shareable across models and frameworks.
To facilitate this, we have built a library containing rules
for {\nkernels} 
standard LLM operators, covering the commonly used models.

%
Fully automated IFT rule generation is not yet implemented in \sys due to the
lack of a complete symbolic execution engine for CUDA programs.
Such an engine would allow \sys to automatically trace all input values that influence specific outputs.
Alternatively, a fully functional transpiler that accurately
converts CUDA programs to C++ would enable \sys to leverage KLEE for this
analysis, but no such tool currently exists.
At present, \sys requires human involvement to manually extract IFGs
and assist in translating CUDA code to C++ (with semi-automated scripts we build).
Reducing this manual effort is our future work.

\subsection{{\asyncIFT}}
\label{ss:async}
\label{sec:multi-stream}

Today's LLM serving frameworks use multiple GPU streams to manage one or more
GPUs, for example, a background stream swaps the KV cache while the main stream
performs computations.
The \sysmonitor creates dedicated CPU threads to handle each GPU stream,
and each thread is responsible for ownership propagation and kernel launch.
These multi-GPU-stream scenarios require additional concurrency control mechanisms on GPU memory accesses for \asyncIFT.

\stitle{Race conditions.}
Multi-stream execution introduces potential race conditions that could
compromise data privacy. Consider a scenario where an attention kernel in the
main stream accesses user A's KV cache while a background stream simultaneously
overwrites this cache with user B's data. Although the IFT marks the attention
computation result as belonging to user A, the actual result contains user B's
private data, constituting a data leak.

To prevent such vulnerabilities, the \sysmonitor must enforce proper
synchronization to eliminate both read-write and write-write races on GPU memory.
A read-write race occurs when one kernel attempts to read memory that
another kernel is actively modifying, while a write-write race involves
concurrent writes to the same memory.

\begin{figure}[htbp]
  \centering
  \includegraphics[width=1.0\columnwidth]{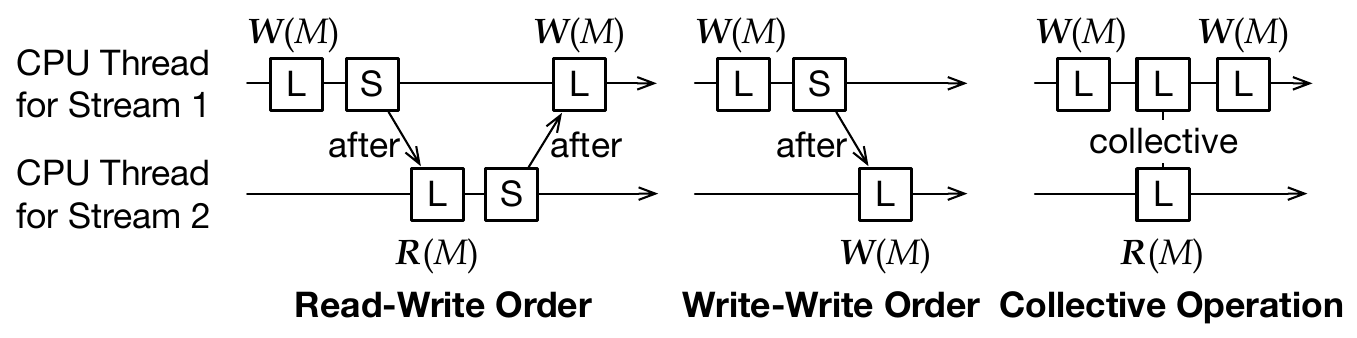}
  \caption{Legal launch order of kernels, L: launch kernel, S: synchronization operation, $R(M)$, $W(M)$: the kernel reads or writes memory segment $M$.}
  \label{fig:concurrency-control}
\end{figure}

\stitle{Concurrency control.}
As illustrated in Figure~\ref{fig:concurrency-control}, the monitor ensures a legal launch order of kernels that access the same memory segment.
For this purpose, it needs to record more information.

The ownership record is extended to track the last writer kernel and the last reader kernels for the memory segment.
Each kernel is represented by a stream ID and its sequence number within that stream.
For each stream, the security monitor also records the last completed kernel.
A kernel is deemed completed if it is launched before a synchronization operation, such as cudaMemcpy or cudaStreamSynchronize.

The monitor prevents read-write races by enforcing two invariants.
First, when launching a kernel that reads memory segment $M$,
if the last writer of $M$ is on another stream, that writer must have completed.
Second, when launching a kernel that writes to memory segment $M$,
if any reader of $M$ is on another stream, that reader must have completed.
Write-write races are prevented in a similar manner.

\stitle{Collective operations.}
For collective operations like all-reduce in tensor parallelism, multiple GPUs run collaborative kernels that exchange data via communication buffers. 
In this scenario, there can be both in-flight reader and writer kernels for the buffer.
Yet, as shown in Figure~\ref{fig:concurrency-control}, the reader on stream 2
(data receiver) does not conflict with the writers on stream 1 as the buffer
$M$ is transferred after the first writer completes, and the transfer finishes
before the second writer starts executing.
The monitor permits this launch order.

\subsection{Token Encryption and KV Cache Optimizations}
\label{sec:encryption}

\stitle{Token encryption.}
Token encryption is needed when users upload their prompts to the serving framework
or \sysmonitor returns the generated tokens to the serving framework.
\sys encrypts each token with a block cipher,
instead of encrypting the entire sequence with a stream cipher,
because the serving framework needs to perform string operations on the sequence,
such as appending newly-generated tokens or splitting prompts into chunks~\cite{sarathi-serve}.

Simple per-token encryption can expose sequence patterns,
as identical tokens yield the same ciphertext with the same user key.
\sys solves this issue with token obfuscation. 
When encrypting a token, the monitor (or users' devices) extends the 64-bit token to 128 bits by appending a random nonce, then processes it through a block cipher.
During decryption, the original token is retrieved by removing the nonce from the 128-bit plaintext.
The monitor creates a special ciphertext for the end-of-sequence token to inform the inference framework that the generation is complete.

\stitle{Secure swap space for KV cache.}
When GPU memory is insufficient, the serving framework swaps the KV cache to CPU memory.
If the KV cache is swapped to the framework's memory, it must be
encrypted to prevent privacy leakage. Encrypting and decrypting a large KV
cache (e.g., 400 KB per token for OPT-13B) incurs significant overhead.

To avoid this overhead, \sysmonitor allocates a secure swap space in its protected memory for the KV cache,
which is inaccessible to the serving framework.
It provides the serving framework with interfaces for data transfer between GPU memory and this swap space, 
and tracks ownership of swapped KV blocks. When blocks are swapped back to GPU memory, their ownership records are restored accordingly.

\stitle{Caching index tables.}
Sometimes the IFT rule depends on the \emph{index tables} of the KV cache in GPU memory.
For example, \sysmonitor needs to access the block table to determine which KV blocks are accessed by the attention kernel. 
However, copying index tables from GPU to CPU memory will block IFT, leading to performance overhead.
\sysmonitor eliminates this overhead by caching a copy of the index tables in its own memory.

\section{\syscc: Isolation with Confidentiality}
\label{sec:cc}

\stitle{Motivation.}
Since LLM serving systems are typically deployed in outsourced environments such as public clouds,
users of these services have a legitimate concern about whether the
underlying systems (e.g., OSes and hypervisors) and the service providers can be fully trusted.
%
Thereby, we design and implement \syscc to tackle a stronger threat model:
how can user data confidentiality be protected when processed in an untrusted LLM serving infrastructure?

\stitle{TEE Background.}
Confidential computing has emerged as a promising solution for protecting sensitive data on cloud platforms.
CPU vendors provide confidential virtual machines (CVMs) like AMD SEV~\cite{amd-sev} and Intel TDX~\cite{intel-tdx};
modern GPUs like NVIDIA H100~\cite{nvidia-cc} now support confidential computing as well.
\syscc is designed to run LLM serving in a heterogeneous (CPU+GPU) trusted execution environment (TEE),
providing user data isolation as well as protecting data confidentiality from the cloud providers.

%


\stitle{Difference from {\sys}.}
\syscc adopts the same overall architecture as \sys, but places its trusted components 
and GPU computation inside TEEs.
Specifically, {\sysmonitor} is deployed in a CVM that exclusively controls the TEE-enabled GPUs.
Before LLM serving, users establish secure communication channels with the monitor using
TEE-provided remote attestation~\cite{nvidia-cc} to confirm \cc is enabled. 

The encryption-as-isolation design ensures that
all data flowing to untrusted components (i.e., those outside the TEEs) are encrypted.
In addition, the CVM communicates with the GPUs through an encrypted PCIe channel,
which enforces exclusive GPU control and prevents physical attacks.
This encryption incurs additional overhead when copying tokens and KV cache (\S\ref{sec:encryption})
between {\sysmonitor} and the GPUs.
\begin{figure*}[htb]
    \centering
    \includegraphics[width=2\columnwidth]{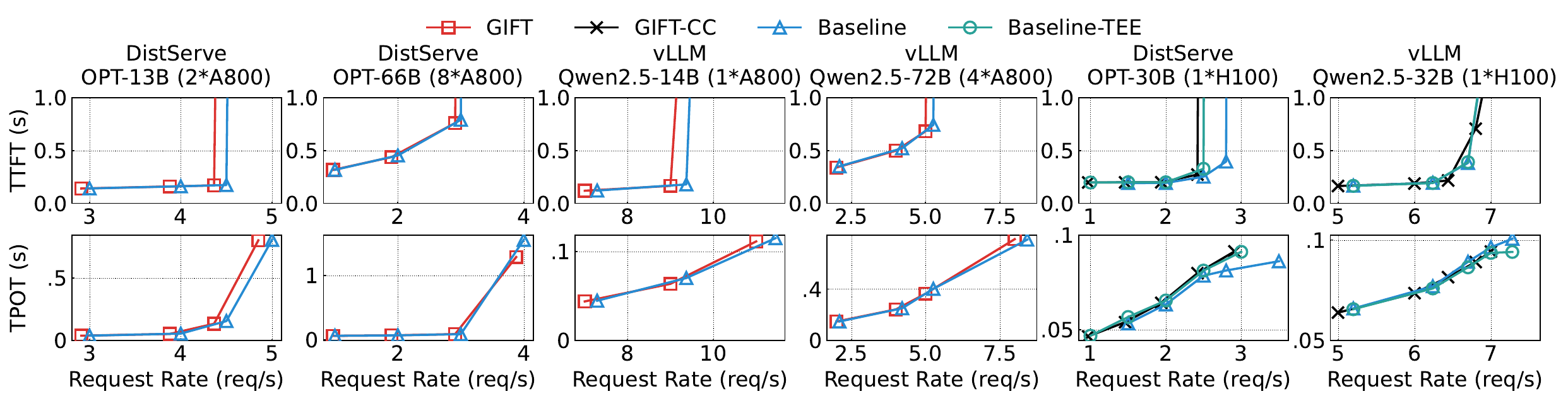}
    \caption{The TTFT and TPOT of ShareGPT benchmark under different request rates on the A800 and H100 machines.}
    \label{fig:latency-eval}
  \end{figure*}
  
  \begin{figure}[htbp]
    \centering
    \includegraphics[width=1\columnwidth]{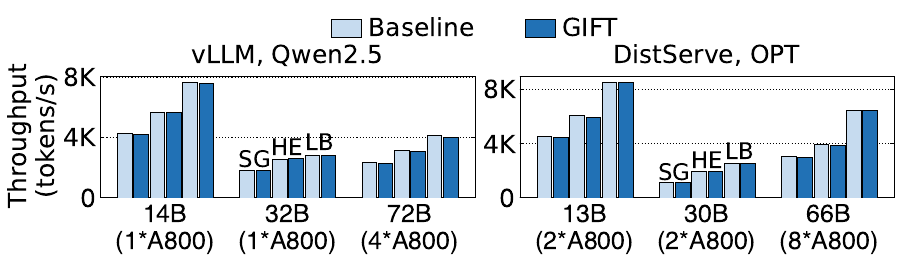}
    \caption{Serving throughput on the A800 machine. SG: ShareGPT, HE: HumanEval, LB: LongBench.}
    \label{fig:a800-eval}
  \end{figure}

\section{Implementation and Limitations}
\label{sec:impl}

\sys's design is general and compatible with existing LLM serving frameworks,
allowing them to freely evolve while enforcing user data isolation.
To demonstrate this,
we implement {\sys} on two representative frameworks: vLLM~\cite{vllm} and
DistServe~\cite{dist-serve}---each with two configurations: with and without \cc.


We implement \sysmonitor in Rust. It maintains a minimal codebase of {\tcb} lines
while exposing only 32 interfaces for kernel registration/launching and memory copying/allocation.
To intercept CUDA API calls with minimal overhead, we adopt and extend optimized techniques for API remoting~\cite{gpu-remoting}.
We modify $<$120 lines in each serving framework to track the request owner and call custom APIs.

We implement a library of IFT rules with {\nkernels} kernel families\footnote{Multiple
kernels that share a similar information flow, such as elementwise-operation
kernels with different functors and data types, are considered as the same kernel type.}
for vLLM and DistServe.
With this library,
\sys works with the models natively supported by these serving frameworks,
including
Qwen-2.5~\cite{qwen2.5-models},
Llama-3~\cite{llama},
Gemma-2~\cite{gemma-2},
OPT~\cite{opt-models},
GPT-2~\cite{gpt-2},
Phi-3~\cite{phi-3}.

\stitle{Limitations.}
\sys currently has three main limitations.
First, it assumes that the CPU-side framework does not modify user data contents.
If this assumption does not hold in some (future) scenarios, \sys requires changes to the framework to
delegate such operations to the \sysmonitor.
Second, some GPU kernels may be unavailable for ahead-of-time analysis, either
due to technical constraints or legal restrictions.
In such cases, users must derive approximate IFT rules through black-box sampling, which may sacrifice soundness.
Third, \sys's current IFT rule generation process is semi-automated, due to the absence of a fully automated CUDA symbolic executor or a comprehensive CUDA transpiler.

\section{Performance Evaluation}
\label{sec:eval}
\label{s:eval}

In this section, we evaluate \sys's performance and defer its security analysis to the next section.
In particular, we mainly answer three questions:
\begin{myitemize}
    \item \emph{End-to-end performance}: How do \sys and \syscc compare to baselines without user data isolation? (\S\ref{ss:eval:e2e})
    \item \emph{Detailed IFT overhead}: What are the costs of tracking information flows for GPU kernels? (\S\ref{ss:eval:cost})
    \item \emph{Optimization impact}: How does each optimization contribute to the overall performance? (\S\ref{ss:eval:opt})
\end{myitemize}

\stitle{Baselines.}
We compare \sys against unmodified vLLM~\cite{vllm} and DistServe~\cite{dist-serve},
referred to as \emph{baselines}.
For \syscc,
we build corresponding TEE-enabled versions that run the original serving frameworks
within a TEE with NVIDIA-CC, which we call \emph{baseline-TEE}.

\stitle{Models.}
For space saving, we present the experimental results of Qwen2.5 models (14B,
32B, 72B)~\cite{qwen2.5-models} on vLLM 
and OPT models (13B, 30B, 66B)~\cite{opt-models} on DistServe. 
The evaluation results are similar for other models. 

\stitle{Framework features.}
We turn on common optimization features
in the serving frameworks. 
Contiguous batching, paged attention, and pipeline/tensor parallelism are enabled for both vLLM and DistServe experiments.
Chunked prefill is supported by vLLM and enabled for its related experiments.
Prefill-decoding disaggregation is enabled for DistServe-related experiments.


\stitle{Benchmarks.}
ShareGPT~\cite{share-gpt} (ChatGPT conversations), HumanEval~\cite{human-eval} (programming tasks), and LongBench~\cite{long-bench} (long text summarization tasks).

\stitle{Testbed.}
The evaluation is conducted on two machines:
(1) a server with 8 NVIDIA A800 GPUs (NVLINK-connected) and Intel Xeon Platinum CPUs (3.4 GHz),
(2) a server with 1 NVIDIA H100 GPU and Intel Xeon Gold CPUs (3.5 GHz).

\emph{Baseline-TEE} is evaluated only on the H100 platform, as A800 does not support NVIDIA-CC.
For \syscc, we modify DistServe to run prefill and decoding stages on the same GPU, as only one H100 GPU is available.
For the same reason (insufficient GPUs), OPT-66B and Qwen2.5-72B are not evaluated for \syscc on the H100 machine.

\subsection{End-to-End Performance}
\label{sec:eval-e2e}
\label{ss:eval:e2e}

We experiment \sys and \syscc with baselines
under different models and workloads.
The performance metrics are
time to first token (TTFT),
time per output token (TPOT),
and system throughput.

\subsubsection{Performance of \sys (on A800)}
\label{sec:a800-eval}

\stitle{Latency.}
Figure~\ref{fig:latency-eval} shows the time to first token (TTFT) and time per output token (TPOT) of ShareGPT benchmark on the A800 machine.
Compared with the baseline, {\sys} incurs up to 5\% overhead on the maximum request rate under the same latency SLO (Service Level Objectives). 
The overhead on other benchmarks is similar. 

\stitle{Throughput.}
Figure~\ref{fig:a800-eval} presents the serving throughput between {\sys} and the baseline on the A800 machine.
We use a high request rate to evaluate the maximum throughput of the systems.
For the DistServe-based system, both {\sys} and the baseline swap the KV cache under this setting,
as the GPU memory is insufficient to accommodate all the KV cache required for a decoding batch.

Compared with the baseline,
{\sys} incurs up to 2.3\% and 4.1\% overhead for the DistServe-based and the vLLM-based systems, respectively.
The overhead is small because the overheads of IFT, encryption, and CUDA API redirection (three main sources that may cause overhead in {\sys}) are minimized.
First, {\sys} hides the IFT overhead with {\asyncift}.
We analyze kernel execution and IFT latency later (\textsection{\ref{sec:eval-ift}}).
We also evaluate the effect of optimizations (\textsection{\ref{sec:eval-opt}}).
%
Second, {\sys} avoids having the security monitor encrypt the swapped KV cache when the CPU memory is sufficient.
We also examine the effect of this optimization in \textsection{\ref{sec:eval-opt}}.
The remaining encryption overhead (related to token encryption) is minor due to the small data size.
Token encryption accounts for less than 1\% of the total serving time for a request.
Third, the API redirection overhead in {\sys} is less than 0.5\%, 
thanks to the shared-memory-based communication.

\subsubsection{Performance of \syscc (on H100)}

\stitle{Latency.}
We evaluate TTFT and TPOT using ShareGPT benchmark on the H100
platform, as illustrated in Figure~\ref{fig:latency-eval} (the two columns on the right).
Under the same latency SLOs, {\syscc} exhibits an overhead on maximum request rate of 7\%
compared to Baseline, and 4\% relative to Baseline-TEE.
Similar overheads are observed across benchmarks.

\begin{figure}[htbp]
  \centering
  \includegraphics[width=1\columnwidth]{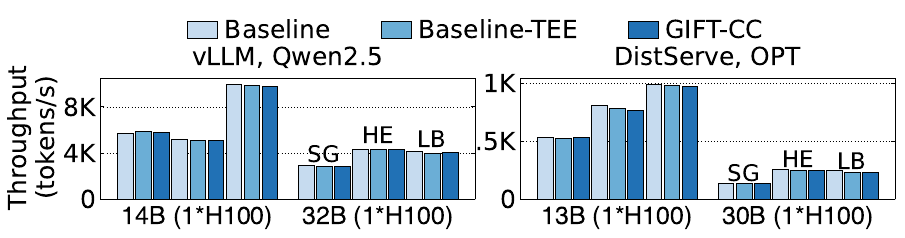}
  \caption{Serving throughput on the H100 machine (TEE enabled), SG: ShareGPT, HE: HumanEval, LB: LongBench.}
  \label{fig:h100-eval}
\end{figure}

\stitle{Throughput.} Figure~\ref{fig:h100-eval} shows the serving throughput of {\syscc}, Baseline, and Baseline-TEE on the H100 machine.
Compared with Baseline, 
{\syscc} brings up to 6.7\% and 3.2\% overhead for the DistServe-based and the vLLM-based systems, respectively.
This additional overhead, relative to the A800 results, stems from the encryption and decryption of data transferred between CPU and GPU memory in the H100 TEE environment.
The observed overhead is moderate due to the limited intensity of KV cache swapping in our experiment setup.
Note that this overhead could be mitigated through emerging TEE hardware architectures~\cite{tdx-connect,pcie-ide}
and orthogonal optimizations like prefetching~\cite{pipellm}.

Compared with Baseline-TEE, {\syscc} brings up to 2.4\% and 1.7\% overhead for the DistServe-based and the vLLM-based systems, respectively.
The overhead is small for the same reasons discussed in \textsection{\ref{sec:a800-eval}}.

\begin{figure}[htbp]
  \centering
  \includegraphics[width=1\columnwidth]{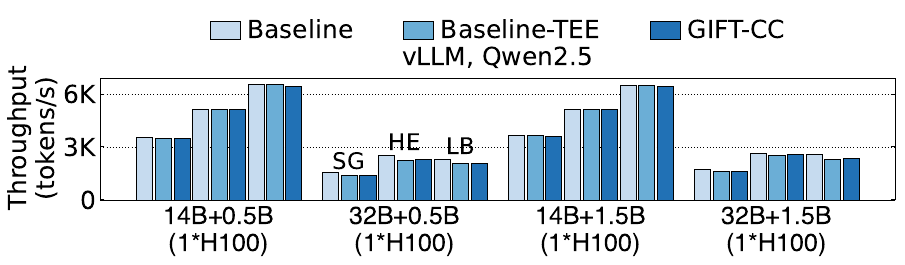}
  \caption{Serving throughput with speculative decoding on the H100 machine (TEE enabled), SG: ShareGPT, HE: HumanEval, LB: LongBench.}
  \label{fig:h100-spec-eval}
\end{figure}

\stitle{Experiments with speculative decoding.}
Figure~\ref{fig:h100-spec-eval} shows the serving throughput of {\syscc}, Baseline, and Baseline-TEE with speculative decoding, evaluated on the H100 machine.
Compared with Baseline, {\syscc} incurs up to 10.7\% overhead due to the CPU-GPU communication overhead of NVIDIA-CC.
Compared with Baseline-TEE, {\syscc} only incurs up to 2.1\% overhead.

\subsection{Costs of Information Flow Tracking}
\label{ss:eval:cost}

In this subsection, we show the costs of IFT for different GPU kernels (\S\ref{sec:eval-ift}),
how their overhead varies with model size and input size (\S\ref{ss:eval:kernelperf}),
and \sys's CPU and memory costs (\S\ref{ss:eval:cpuoverhead}).

\subsubsection{GPU Kernel IFT Performance}
\label{sec:eval-ift}

\begin{figure}[htbp]
  \centering
  \includegraphics[width=1\columnwidth]{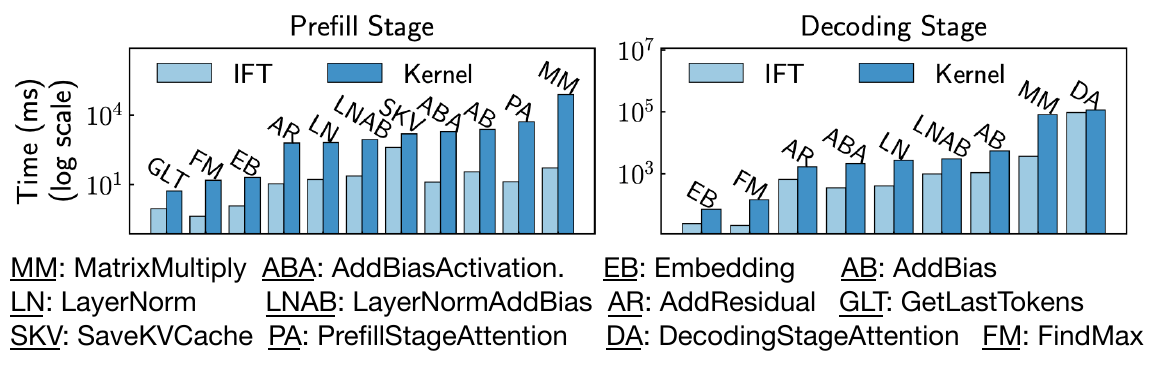}
  \caption{The total execution time and IFT time for each GPU kernel (DistServe, ShareGPT, OPT-13B, A800).}
  \label{fig:kernel-ift-duration}
\end{figure}

Figure~\ref{fig:kernel-ift-duration} shows the total execution time and IFT time of each kernel in the DistServe-based system when running the ShareGPT benchmark with OPT-13B on the A800 machine.
For every kernel, the IFT time is significantly lower than the kernel execution time (note that the Y-axis is log-scale).

Therefore, the IFT overhead can be hidden with {\asyncift}.
Specifically, the framework iteratively launches kernels for each model layer,
and thus the IFT for one kernel can overlap with the execution of previously-launched kernels.
For instance, we observe that the IFT time of a DecodingStageAttention kernel can be hidden within the execution time of a previous MatrixMultiply kernel.

\subsubsection{IFT Overhead Analysis}
\label{ss:eval:kernelperf}

\stitle{IFT overhead under different model parameter sizes.}
The parameter size of one kernel depends on the model's hidden size and the number of GPUs used for tensor parallelism.

\begin{figure}[htbp]
  \centering
  \includegraphics[width=1\columnwidth]{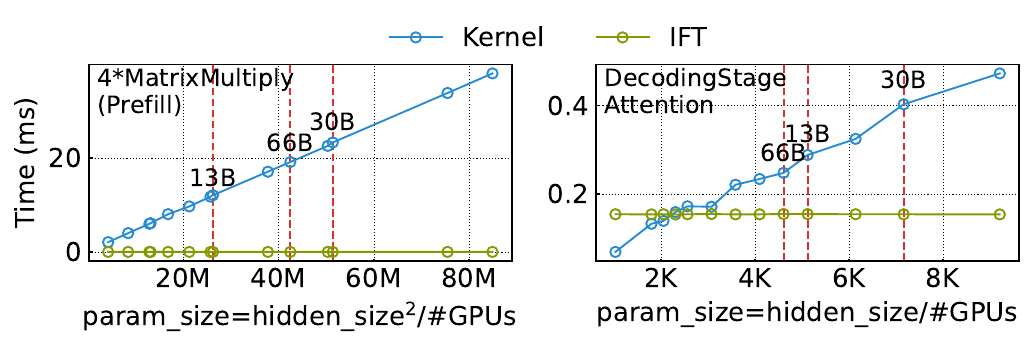}
  \caption{Kernel execution and IFT time under different parameter sizes (A800), \#GPUs: tensor parallelism group size.}
  \label{fig:kernel-ift-latency-param}
\end{figure}

Figure~\ref{fig:kernel-ift-latency-param} shows the execution time and IFT time under various parameter sizes for two representative kernels with short or long IFT time.
The red lines show the setups used in \textsection{\ref{sec:a800-eval}} (OPT models).
As the parameter size increases, 
the kernel execution time rises, while the IFT time remains constant.
The IFT time is constant because the ownership of a hidden state tensor or a KV block can be read or written in a single operation, regardless of their sizes.
For the DecodingStageAttention kernel, the IFT time exceeds the execution time when the parameter size is smaller than 2,048,
which can be further reduced via multi-threaded IFT (future work).



\stitle{IFT overhead under different input sizes.}
The amount of input data involved in kernel execution depends on the batch size (the number of requests) and the context length of each request.
Context length refers to the prompt length in prefill stage and the number of previous tokens in decoding stage.

\begin{figure}[htbp]
  \centering
  \includegraphics[width=1\columnwidth]{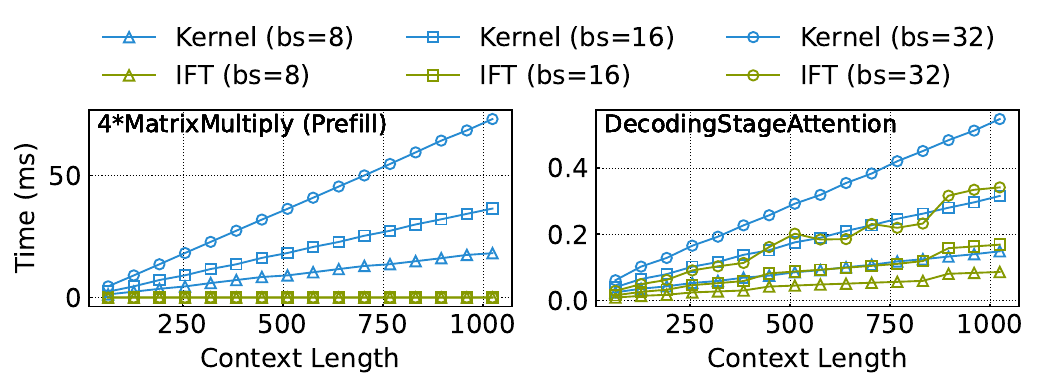}
  \caption{Kernel execution and IFT time under different input sizes (OPT-13B, A800), bs: batch size (\#requests).}
  \label{fig:kernel-ift-latency}
\end{figure}

In prefill stage, the kernel execution time grows with larger batch size or context length. Figure~\ref{fig:kernel-ift-latency} illustrates the execution time for MatrixMultiply under different input sizes.
The IFT time remains low because it only depends on the batch size, as the hidden states of all tokens in a single prompt share the same owner. 
In contrast, for most kernels (except for the attention kernel) in decoding stage,
the kernel input size is only related to the batch size.
Therefore, the gap between the kernel execution time and its IFT time 
is smaller in decoding stage than in prefill stage (Figure~\ref{fig:kernel-ift-duration}).

Among all the kernels, DecodingStageAttention has the longest IFT time.
Figure~\ref{fig:kernel-ift-latency} shows that both its execution time and IFT time increase with batch size and context length,
because it calculates attention score on each previous token, 
and its IFT traverses each KV block accessed by the kernel.
The IFT time grows slower than the execution time because ownership transfer is much simpler than attention calculation.
Across all different input sizes, the IFT time remains lower than the execution time.

\subsubsection{CPU and Memory Overhead}
\label{ss:eval:cpuoverhead}

Compared with Baseline, 
{\sys} incurs CPU overhead due to IFT, encryption, and API redirection,
and needs extra CPU-side memory for recording ownership.
Across all setups on both the A800 and H100 machines, 
the additional CPU usage is below 124\% (1.24 cores),
and the memory overhead is under 130 MB owing to {\batchedIFT}.



\subsection{Effects of Performance Optimizations}
\label{sec:eval-opt}
\label{ss:eval:opt}

In this subsection, we analyze the effects of \sys techniques
for reducing the overhead of IFT and KV cache swapping,
based on an evaluation on the DistServe-based system (ShareGPT, OPT-13B, 1$\times$A800).

\stitle{Fine-grained concurrency control.}
A naive concurrency control blocks all but one GPU stream when races may occur, which will block the prefill GPU when the decoding GPU fetches KV cache.
Our fine-grained concurrency control (\textsection{\ref{sec:multi-stream}}) avoids unnecessary blocking, enabling more concurrency and thus improving {\sys}'s throughput by 12\%.

\stitle{Secure swap space for KV cache.}
{\sysmonitor} saves swapped KV blocks in its protected memory,
which avoids the encryption and decryption overhead of copying KV cache between GPU and the serving framework (\textsection{\ref{sec:encryption}}).
Without this optimization, {\sys} experiences an 11\% reduction in serving throughput compared with Baseline, while the optimization reduces the overhead to 1.9\%.

\stitle{Caching index tables.}
{\sysmonitor} caches index tables used by GPU kernels in the CPU memory (\textsection{\ref{sec:encryption}}).
In our experiments, there are four kernels, including DecodingStageAttention, that utilize index tables. 
Without caching, the monitor needs to load the index table from GPU memory
when performing IFT for these kernels, which will block until the previously-launched kernels complete and thus expose the IFT overhead on the critical path.
The caching optimization increases the throughput of {\sys} by 50\%.

\section{Security Analysis}
\label{sec:security-analysis}

{\sys} ensures user data isolation even when the LLM serving framework is compromised.
Furthermore, {\syscc} protects user data confidentiality against attackers with administrative privileges.

\stitle{Preventing direct access attacks.}
If the serving framework (e.g., vLLM) is compromised and an attacker attempts
to read user data from CPU memory, the attacker encounters only encrypted
content protected by user-specific keys.
If the attacker tries to access plaintext data in either the \sysmonitor memory or GPU memory,
such attempts are blocked,
by VM isolation in the case of \sys (which assumes a trusted OS/hypervisor),
or by the hardware-enforced TEE boundary in the case of \syscc.

\stitle{Preventing confused-deputy attacks.}
If an attacker attempts to launch a malicious GPU kernel that mixes data from
different users (e.g., combining the victim's data with the attacker's) in GPU
memory, \sysmonitor rejects the kernel according to the generated IFT rules.
Even if the attacker issues a sequence of GPU kernels to propagate the victim's
data across memory regions before extraction,
\sysmonitor continuously tracks memory ownership and ensures that the correct encryption key is applied,
preventing the attacker from accessing private user data.


\stitle{Preventing physical attacks by \syscc.}
For physical attack vectors, {\syscc} relies on the security features of
NVIDIA-CC hardware. The CPU-side TEE (CVM) employs memory encryption to protect against
memory inspection, while the GPU HBM is resistant to common physical attack
tools~\cite{nvidia-cc}. Additionally, NVIDIA-CC secures PCIe transactions between GPUs and CPUs
through TEE-enforced encryption, and extends this protection to NVLINK channels.

\stitle{TCB Analysis.}
{\syscc}'s TCB includes TEE hardware and the software in the TEE: a minimal Linux kernel (1M LoC) as the CVM (CPU TEE) OS, CUDA runtime (unknown LoC), CUDA driver (1.4M LoC), and our security monitor ({\tcb} LoC in Rust). 
While this may seem substantial, {\syscc} significantly reduces the overall TCB by excluding the LLM serving  framework and its dependencies (3.6M LoC for vLLM) and removes the full-fledged host Linux and other cloud infrastructure (estimated at over 30M LoC) from the TCB.

The remaining TCB, though not minimal, offers strong security for two reasons: (1) it presents a limited attack surface with only 32 exposed interfaces, and (2) it comprises relatively steady and secure components—notably, CUDA has maintained a steady codebase with no public CVEs reported in the past decade~\cite{cvedetails}. The minimal Linux kernel in the CVM is specifically tailored to remove vulnerable drivers and management interfaces. Moreover, compared to the standard security model of NVIDIA-CC (where CUDA is designed to be in the TCB), {\syscc} adds only $\sim${\tcb} lines of Rust code to the TCB.

Without TEE, {\sys} needs to rely on the underlying hypervisor to provide an isolated VM as a protected domain for \sysmonitor. Thereby, the TCB of {\sys} further includes the hypervisor. While current {\sys} implementation uses vanilla KVM as the hypervisor, existing secure hypervisors~\cite{9519433,BlackBox,InkTag,CloudVisor} could be integrated to reduce the TCB.

\section{Related Work}
\label{sec:related-work}

\stitle{Information flow tracking (IFT).}
IFT is a general method for tracking data flow at runtime, primarily in CPU programs.
It can help detect information leakage, identify abnormal behavior, etc~\cite{taint-analysis,taint-droid,data-lifetime,dytan,demand-emulation,lift,taint-enhanced-policy,taint-eraser,Ryoan}.
To reduce the overhead of IFT, prior work~\cite{shadow-replica,speck,cab,pipa} has explored offloading flow tracking to a separate CPU thread. While our \asyncift shares a similar spirit, it is specifically designed for and applied within the GPU context.
The initial effort of implementing
GPU taint tracking~\cite{gpu-taint} instruments GPU kernels to monitor the data flow in GPU memory.
It incurs {2.5--5.7\texttimes} slowdown on application performance.
Compared to prior work, {\sys} is the first system to address the unique
challenges of implementing IFT for GPU-based LLM serving, achieving near-zero runtime overhead.


\stitle{GPU kernel analysis.}
\sys analyzes GPU kernels to generate IFT rules.
Prior work has applied static or dynamic analysis to GPU kernels~\cite{DBLP:conf/nfm/ChiangGLR13,DBLP:conf/sigsoft/LiG10,DBLP:journals/toplas/BettsCDKQTW15,DBLP:conf/oopsla/BettsCDQT12,DBLP:conf/sosp/KamathB21,DBLP:conf/pldi/LeungGAGJL12,DBLP:conf/ppopp/LiLSGGR12,pos,honeycomb},
but none addresses ownership tracking for GPU IFT.
For example, POS~\cite{pos} and Honeycomb~\cite{honeycomb} analyze memory accesses in GPU kernels for fault tolerance or illegal access prevention, 
whereas \sys constructs an information flow graph to enable ownership tracking. 

\stitle{Privacy-protected LLM serving.}
Cryptography-based solutions such as Homomorphic
Encryption~\cite{Iron-NIPS22,zimerman2023convertingtransformerspolynomialform,chen2022thexprivacypreservingtransformerinference},
Multi-Party Computation~\cite{9804616,dong2023pumasecureinferencellama7b},
Functional Secret Sharing~\cite{gupta2023sigma}, and Differential
Privacy~\cite{DP-Forward-CCS23} can secure user privacy in outsourced LLM serving.
However, the significant performance overhead and model modifications that can affect accuracy hinder their adoption.
In addition, current studies are primarily evaluated on relatively small models (e.g., 7B).

Protecting AI applications within TEE has been extensively studied~\cite{strongbox,snpu,telekine,graviton,hix,gevisor,cage,acai,hetee,honeycomb,cure}.
Recent studies on running LLMs inside TEEs~\cite{10643934,pipellm}
show that the performance of TEE-based LLM serving can be close to the non-confidential baseline.
\sys is orthogonal to these systems and focuses on enforcing user data isolation.
By combining the two, we implement \syscc,
which protects user privacy in a shared LLM inference service.

Apple uses cloud-based LLMs to enhance the user experience of end devices~\cite{apple-intelligence}.
Apple Private Cloud Compute (PCC)~\cite{pcc} aims to protect privacy for this scenario,
employing techniques like OS/libraries tailoring and hardening, and minimizing management interfaces to achieve a narrow attack surface. {\syscc} shares the same security goal as PCC, but differs in the approach.
First, PCC is specific to Apple's proprietary hardware and software,
while {\syscc} is more portable.
Second, PCC requires users to trust its entire infrastructure including the inference framework,
which necessitates a larger TCB than {\syscc}.



\section{Conclusion}
\label{sec:conclusion}

{\sys} is the first IFT design for LLM serving systems, which enables user data isolation.
Through three key designs, {\batchedIFT}, {\whiteboxIFT}, and {\asyncIFT}, 
{\sys} introduces minimal performance overhead compared with vanilla LLM serving systems.
Furthermore, by integrating confidential computing, {\syscc} offers stronger privacy protection against untrusted cloud environments.


\balance

\small{
\bibliographystyle{abbrv}
\bibliography{paper}
}

\end{document}